# Defining the relationship between cathepsin B and esophageal adenocarcinoma: conjoint multi-omics analysis

Jialin Li[1], Shaokang Yang[2], Xinliang Gao[1], Mingbo Tang[1], Xiaobo Ma[3], Suyan Tian[4*], Wei Liu[1*]

[1] Department of Thoracic Surgery, The First Hospital of Jilin University, 1 Xinmin Street, Changchun, Jilin 130021, P.R. China

[2] Department of Gastric and Colorectal Surgery, General Surgery Center, The First Hospital of Jilin University, 1 Xinmin Street, Changchun, Jilin 130021, P.R. China

[3] Department of Pathology, The First Hospital of Jilin University, 1 Xinmin Street, Changchun, Jilin 130021, P.R. China

[4] Department of Clinical Research, The First Hospital of Jilin University, 1 Xinmin Street, Changchun, Jilin 130021, P.R. China

*Corresponding authors: Suyan Tian. Address: The First Hospital of Jilin University, 1 Xinmin Street, Changchun, Jilin 130021, P.R. China. Email: wmxt@jlu.edu.cn; Wei Liu. Address: The First Hospital of Jilin University, 1 Xinmin Street, Changchun, Jilin 130021, P.R. China. Email: l_w01@jlu.edu.cn.

**Abstract**

Esophageal cancer poses a significant global health challenge, with the incidence of esophageal adenocarcinoma (EAC), a predominant subtype, increasing notably in Western countries. Cathepsins, a family of lysosomal proteolytic enzymes, have been implicated in the progression of various tumors. However, the causal relationship between the cathepsin family and EAC remains unresolved. To evaluate these potential causal associations, integrative analyses were conducted, integrating Mendelian randomization (MR), TWAS, scRNA-seq, and sc-eQTL analyses. MR analyses demonstrated elevated levels of cathepsin B (CTSB) were associated with a reduced risk of EAC. The TWAS analysis identified a negative association between CTSB expression in esophageal tissue and EAC, consistent with experimental validation using immunohistochemistry. The scRNA-seq analysis indicated CTSB expression was predominantly localized in macrophages infiltrating EAC. Colocalization analysis incorporating sc-eQTL data specific to macrophages confirmed a shared causal variant between CTSB and macrophages. Additionally, MR analysis of CTSB and macrophage scavenger receptor (MSR) established their interrelationship, suggesting CTSB may influence the proinflammatory phenotype of macrophages, ultimately affecting EAC risk. This integrative analysis identified a significant causal association between CTSB and EAC, potentially mediated through macrophage MSR regulation. These findings suggest targeting cathepsin B could represent a novel strategy for the diagnosis and treatment of EAC.

# Introduction

Esophageal cancer is a significant global health issue and ranks as the sixth leading cause of cancer-related mortality [1]. It is classified into two primary subtypes: esophageal squamous cell carcinoma (ESCC) and esophageal adenocarcinoma (EAC) [2]. Over recent decades, the incidence of EAC has increased more rapidly than any other cancer, particularly in developed nations [3]. Due to the lack of effective and noninvasive early screening methods, most patients are diagnosed at advanced stages, resulting in a 5-year survival rate of less than 20% [4]. EAC generally develops from Barrett's mucosa in the lower esophagus, characterized by intestinal metaplasia, which leads to Barrett's esophagus (BE), a precursor to EAC [5]. While EAC is the most common pathological type in Western countries, the progression from reflux esophagitis to BE, dysplasia, and adenocarcinoma is not fully understood under the Western model of EAC formation [6].

Cathepsins, a family of lysosomal proteolytic enzymes, play a role in the autophagy-lysosome and ubiquitin-conjugating pathways, facilitating proteolytic degradation in lysosomes [7]. Dysregulated expression or activity of cathepsins contributes to the pathogenesis of cancer, neurodegeneration, and autoimmune diseases [8]. Previous research has shown that the autophagy-lysosome proteolytic system and enzymatic activities are altered in esophageal cancer [9]. Consequently, studies have explored the link between cathepsins and esophageal cancer. Elevated levels of cathepsin D (CTSD) have been observed in EAC cell lines [10]. Increased cathepsin E (CTSE) mRNA expression in EAC tissues has been associated with a decreased risk of mortality [11]. Additionally, aberrant activities of cathepsin B (CTSB), C (CTSC), and S (CTSS) have been reported in EAC/BE tissues [12,13]. However, evidence from prospective observational studies and clinical trials examining the causal relationships between specific cathepsins and EAC remains limited.

Mendelian randomization (MR) is an epidemiologic approach for investigating the causal effects of exposures on outcomes by leveraging genetic variants identified through genome-wide association studies (GWAS). Transcriptome-wide association studies (TWAS) provide insights into gene expression levels, facilitating the identification of susceptibility genes for complex diseases. Additionally, single-cell RNA sequencing (scRNA-seq) is a powerful method for elucidating the microscopic pathogenic mechanisms of risk factors at the single-cell level. Single-cell expression quantitative trait locus (sc-eQTL) studies further identify genetic variants specific to particular cell subsets. This study integrated MR, TWAS, scRNA-seq, and sc-eQTL analyses to comprehensively assess the impact of cathepsins on EAC risk, considering plasma protein, cellular, genetic, and gene expression levels.

**Results**

**Defining the causal relationships between various cathepsins and EAC/BE**

The causal relationships between nine cathepsins (B, E, F, G, H, L2, O, S, and Z) and EAC/BE were assessed through two-sample univariable MR analysis. The IVW method indicated that CTSB was significantly associated with EAC and BE risk (p = 0.001, odds ratio [OR] = 0.890, 95% confidence interval [CI] = 0.833–0.951). No statistically significant causal associations were observed for the other eight cathepsins (Figure 1). MR-Egger intercept tests showed no evidence of directional pleiotropy, and MR-PRESSO tests identified no outliers in the univariable MR analyses of cathepsins with EAC/BE risk. The detailed results are presented in Supplementary Table 1.

Subgroup analyses were conducted for EAC and BE. In patients with EAC, univariable MR analysis revealed that high CTSB expression levels were significantly associated with a reduced risk of EAC (p = 0.004, OR = 0.866, 95% CI = 0.786–0.954), and cathepsin S showed a suggestive association with EAC risk (p = 0.007, OR = 0.897, 95% CI = 0.829–0.971). For BE, MR analysis of cathepsins and BE revealed a suggestive causal relationship between CTSB and BE (p = 0.017, OR = 0.910, 95% CI = 0.843–0.983), cathepsin Z and BE (p = 0.030, OR = 1.086, 95% CI = 1.008–1.170) (Figure 1). The MR-Egger intercept and MR-PRESSO global tests indicated no evidence of horizontal pleiotropy or outliers (Supplementary Table 1). Reverse MR analysis was performed (Supplementary Table 2), and no reverse causality was observed between any of the cathepsin types and EAC/BE, including subgroup analyses for EAC and BE.

Multivariable MR analysis examined the genetic predisposition for multiple cathepsin types concerning EAC/BE risk. Even after adjusting for other cathepsins, CTSB levels remained inversely associated with EAC/BE risk (p = 0.004, OR = 0.896, 95% CI = 0.831–0.966). Similar associations were observed in separate analyses for EAC (p = 0.034, OR = 0.899, 95% CI = 0.814–0.992) and BE (p = 0.019, OR = 0.899, 95% CI = 0.822–0.982) (Figure 2). However, after adjustment for other cathepsin types, no significant causal associations were identified between cathepsin S and EAC, cathepsin Z and BE, or other cathepsins and EAC/BE.

In conclusion, findings from univariable MR, multivariable MR, and sensitivity analyses consistently indicated that high CTSB expression levels are associated with a reduced risk of EAC. Therefore, CTSB was prioritized for subsequent investigations to elucidate the underlying mechanisms influencing its role in EAC risk reduction.

**Defining the relationship between protein-coding gene expression and EAC/BE**

Cathepsin B is encoded by the CTSB gene; therefore, the expression profile of CTSB gene was analyzed to investigate the potential causal relationship between cathepsin B and EAC/BE at the genetic level. The expression levels of the CTSB gene in EAC/BE were assessed by integrating genome-wide association study (GWAS) data for EAC/BE with transcriptome-wide association study (TWAS) data derived from esophageal tissue. Using the TWAS/FUSION method, a negative association between CTSB gene expression in esophageal tissue and the risk of EAC/BE was identified. These results aligned with plasma protein-level analyses (Table 1), indicating that higher CTSB gene expression levels were associated with a lower risk of EAC/BE.

The analyses confirmed that CTSB levels, both at the protein and gene expression levels, are inversely associated with the risk of EAC. To experimentally validate these findings, IHC staining was performed on eight pairs of cancerous and normal esophageal tissues (with normal tissues taken more than 5 cm from the tumor margins). The results (Figure 3a) demonstrated that CTSB levels were higher in paired normal esophageal tissues compared to tumor tissues. A paired-samples Wilcoxon signed rank test yielded a p-value of 0.016 for the proportion of positively stained areas (Figure 3b), supporting the conclusion that higher CTSB levels are associated with a reduced EAC risk.

**Defining the CTSB-related cell subpopulation**

To investigate the pathways or molecular mechanisms influencing the onset and progression of EAC at the single-cell level, scRNA-seq data from EAC patients were analyzed to determine cell types and proportions. After filtering out low-quality cells, 43,349 cells were retained. UMAP plots were generated to visualize the cells in two-dimensional space. Based on canonical markers, eight distinct cell subtypes were identified: T lymphocytes, B lymphocytes, epithelial cells, fibroblasts, myeloid cells, endothelial cells, mast cells, and proliferative cells (Figure 4a). The signature genes and canonical markers used for cluster annotation are presented in Figure 4b. UMAP analysis indicated that epithelial cells and lymphocytes constituted the major cell subpopulations in EAC. To determine CTSB-related target cells, the distribution of CTSB gene expression was visualized on a UMAP plot, showing high expression in the myeloid subpopulation (Figure 4c).

To precisely identify the specific CTSB-associated cell subpopulation, further clustering of the myeloid cells was performed. This revealed three distinct subtypes: macrophages, monocytes, and dendritic cells (DCs), as shown in Figure 4d. Canonical markers used for annotation are presented in Figure 4e, with macrophages identified as the predominant subpopulation, followed by monocytes. UMAP mapping of CTSB distribution across myeloid cells highlighted macrophages as the primary CTSB-associated target cells (Figure 4f). To assess whether CTSB expression in macrophages from EAC patients influences tumor progression, a Kruskal-Wallis test was conducted to compare CTSB expression values across various tumor (T) stages. The analysis results (Supplementary Figure 1) demonstrated that CTSB expression levels tended to decrease with advancing T stages ($p = 0.011$), with significant differences observed in the pairwise comparison between T4 and T3 stages.

**Defining the interaction between cathepsin B and macrophages**

The analyses revealed that CTSB was predominantly expressed in macrophages infiltrating EAC, as shown by the scRNA-seq analysis. To further evaluate the role of macrophages in this relationship, colocalization analysis was performed. A shared locus, rs1692811, was identified between cathepsin B and macrophages. This locus is located at 8p23.1 (11845539) within the CTSB coding region (8p23.1 11842524–11868087). According to macrophage sc-eQTL data, rs1692811 affects the expression levels of CTSB, FDFT1, and LONRF1, with p-values of $3.28 \times 10^{-6}$, 0.007, and $8.98 \times 10^{-5}$, respectively. These findings suggest that rs1692811 regulates CTSB expression in macrophages, potentially contributing to a decreased risk of EAC.

Next, the interactions between macrophages expressing CTSB and other cell populations in EAC tissue were assessed using cell communication analysis. Macrophage subpopulations were stratified into CTSB+ and CTSB- groups based on CTSB expression levels. The UMAP plot and dot plot for annotation of these subpopulations are provided in Supplementary Figure 2. Figure 5a illustrates the communication patterns and corresponding strengths among various cell subpopulations in EAC, while Figure 5b summarizes the outgoing and incoming signaling pathways of these interactions. Results showed that both outgoing and incoming signals were stronger for CTSB+ macrophages compared to CTSB- macrophages. Interaction counts and strengths between CTSB+/CTSB- macrophages and other cell subpopulations are displayed in Figures 5c and 5d. Compared to CTSB- macrophages, CTSB+ macrophages exhibited more frequent and stronger interactions, particularly with DCs, endothelial cells, and monocytes. They also established connections with mast cells, T lymphocytes, proliferative cells, and fibroblasts. Ligand-receptor pair analyses further supported these findings, with dot plots (Figure 5e) showing a greater number of ligand-receptor pairs between CTSB+ macrophages and DCs, endothelial cells, and monocytes. In summary, CTSB+ macrophages engage in more extensive interactions with other cell subpopulations in EAC tissue, potentially influencing the antitumor immune responses of EAC patients. These findings highlight the critical role of CTSB+ macrophages play in the cellular microenvironment of EAC.

**MR analyses exploring the relationship between cathepsin B and macrophages**

As discussed, CTSB is highly expressed in macrophages in EAC patients. Macrophages, which serve as a bridge between innate and adaptive immunity, exhibit complex phenotypes and functions [14]. To further explore the interaction between cathepsin B and macrophages in EAC, MR analysis was conducted to evaluate the causal associations between cathepsin B and various macrophage receptor types, which are essential for macrophage functions [15]. The results of two-sample MR analysis between CTSB and macrophage receptors are presented in Supplementary Table 3. Using the IVW method, a significant association was identified (p = 0.004, OR = 1.122, 95% CI = 1.038–1.213). This result indicated CTSB influenced the expression of macrophage scavenger receptor types I and II (class A receptors, known as MSR-A), which play a role in antigen presentation [16], [17]. Reverse MR analysis revealed that the expression of macrophage scavenger receptor types I and II influenced CTSB levels (IVW method: p = $5.845 \times 10^{-5}$, OR = 1.285, 95% CI = 1.137–1.452). The MR-Egger intercept and MR-PRESSO global tests showed no evidence of directional horizontal pleiotropy (p > 0.05, Supplementary Table 3).

In contrast, no causal association was observed between CTSB and other macrophage receptors, such as the macrophage mannose receptor (CD206), a marker for M2 macrophages [15] (IVW method: p = 0.205, OR = 1.052, 95% CI = 0.973–1.137; reverse analysis: OR = 0.984, p = 0.692, 95% CI = 0.911–1.064). In summary, CTSB affects the expression of macrophage scavenger receptor types I and II, while upregulation of these receptors may lead to increased CTSB expression. This positive feedback loop between CTSB and macrophage scavenger receptor types I and II could play a role in the development and progression of EAC via potentially influencing macrophage phenotype.

Using the ZDOCK protein docking program, potential structures of the MSR-CTSB complex were predicted, and the ZDOCK score was calculated. The pbd_ID values of CTSB and MSR were 2IPP and 7DPX, respectively. The top 10 ZDOCK scores for the MSR-CTSB complex were as follows: 1238.918,

1221.663, 1188.465, 1178.531, 1175.501, 1162.458, 1156.746, 1153.236, 1138.205, and 1121.395. A ZDOCK score exceeding 1000 indicates strong molecular interactions, with higher scores suggesting stronger docking capability. The optimal docking model, shown in Figure 6, revealed that six hydrogen bonds were formed within the CTSB-MSR complex. This strong docking capability suggests that a direct interaction may exist between CTSB and MSR, and CTSB may play a role in the post-transcriptional regulation of MSR.

**Discussion**

The incidence of EAC has significantly risen, especially in Western countries [18]. In addition to gastroesophageal reflux and obesity, genetic susceptibility and associated gene variants contribute to EAC pathogenesis by influencing tumor-related inflammation [19], DNA damage and repair [20], and metabolic processes [21]. In this study, univariable and multivariable MR analyses demonstrated that elevated levels of CTSB are linked to a reduced risk of EAC/BE. This finding aligns with the lower CTSB expression levels observed in EAC tissue through TWAS analysis and was further validated experimentally via IHC. The scRNA-seq analysis revealed significant enrichment of CTSB in macrophages within EAC tissues, while colocalization analysis based on macrophage sc-eQTLs identified rs1692811 as a shared causal variant between these traits. Additionally, cell communication analysis showed that CTSB+ macrophages established more extensive interactions with other cell subpopulations in EAC compared to CTSB- macrophages.

Experimental and epidemiological studies suggest that CTSB exhibits both cancer-promoting and cancer-inhibiting effects depending on the tumor type [22,23]. A recent study highlighted the proteolytic activity of CTSB in the extracellular matrix, implicating its role in tumor invasion and metastasis [24].

Conversely, other research has demonstrated that CTSB induces lysosomal membrane permeabilization, triggering cathepsin-mediated cancer cell death, a critical tumor-suppressor mechanism [25,26]. For EAC specifically, Ali *et al.* [27] reported that the risk variant at chr8p23.1 in EAC cell lines is linked to multiple gene targets, including CTSB. This study provides some evidence that CTSB predominantly exerts an anticancer effect in EAC/BE through cathepsin-mediated cancer cell death, rather than its extracellular matrix proteolytic activity, underscoring its tumor-suppressive function in EAC.

The scRNA-seq and colocalization analyses revealed that changes in the expression of cathepsin B regulated by rs1692811 in patients with EAC/BE were primarily observed in infiltrating macrophages. Further MR analysis demonstrated a mutual relationship between CTSB levels and macrophage scavenger receptor types I and II, indicating that CTSB expression in macrophages is associated with the upregulation of MSR, located in the same region (8p22: 16107881–16192651). As a pattern recognition receptor, MSR is critical for phagocytosis [28] and the release of proinflammatory cytokines in macrophages [29], thereby playing a crucial role in promoting immune responses in EAC patients. No causal relationship was observed between CTSB and the macrophage mannose receptor (CD206), a marker of M2 macrophages, which is associated with immunosuppressive activity. These findings suggest the presence of a potential positive feedback mechanism between CTSB and the proinflammatory phenotype of macrophages, as opposed to the immunosuppressive phenotype. This feedback may exert a significant inhibitory effect on the occurrence and progression of EAC/BE.

This study has several limitations. First, due to the limited sample size of the exposure GWAS data used in the MR analysis, a more lenient significance threshold ($5 \times 10^{-6}$) was applied for genetic associations. While this allowed for the inclusion of more SNPs, it potentially increased the risk of weak instrumental variables and horizontal pleiotropy. Additionally, only GWAS summary data for nine cathepsins were available from public repositories, restricting the ability to investigate causal relationships between other cathepsins previously reported in the literature and EAC. Second, the study primarily included individuals of European ancestry. As a result, the generalizability of the findings to other populations remains uncertain and requires further investigation. Third, partial overlap between macrophage-related GWAS data and cathepsin-related GWAS data may have introduced bias in the estimated effect size, potentially leading to results more reflective of observational studies. Furthermore, while strong docking between CTSB and MSR was identified using the ZDOCK model, the possibility that CTSB regulates MSR expression through other pathways cannot be excluded. Further mechanistic studies are needed to clarify the relationship between CTSB and macrophage phenotypes. Finally, most findings in this study were derived from bioinformatics analyses. The only experimental validation involved a small sample size, which assessed whether CTSB expression was downregulated in tumor tissues. Larger-scale experimental studies involving more subjects are necessary to confirm these findings. Addressing these limitations will be a key focus of our future research efforts.

Overall, this comprehensive analysis of multi-omics data demonstrated a significant causal relationship between CTSB and EAC through MR, TWAS, scRNA-seq, and sc-eQTL analyses. These findings were consistent with the IHC experimental results. The causal effect may be mediated through EAC-infiltrated MSR regulation. These results suggest that CTSB could serve as a potential target for EAC intervention and treatment, offering valuable insights for advancing the diagnosis and therapeutic strategies for EAC.

**Methods**

**Experimental Data**

The GWAS data for cathepsins were obtained from the INTERVAL study, which comprised 3,301 participants from England [30]. Summary data for macrophages were sourced from the Open GWAS project (https://gwas.mrcieu.ac.uk), which included 8,243 individuals of European descent [31]. Additionally, summary data for EAC and BE were retrieved from the GWAS Catalog (https://www.ebi.ac.uk/gwas/), encompassing 6,167 patients with BE, 4,112 individuals with EAC, and 17,159 controls from Europe, North America, and Australia [32].

Germline genotype and gene expression data for esophageal tissues were sourced from the Gene-Tissue Expression project (GTEx; version 8) [33]. scRNA-seq data for EAC were obtained from the Gene Expression Omnibus (www.ncbi.nlm.nig/gov/geo) under accession number GSE173950. Macrophage-related sc-eQTL data were derived from the sc-eQTL database (http://bioinfo.szbl.ac.cn/scQTLbase/). All studies underwent review and received approval from institutional ethics review boards at the respective institutions involved.

**MR analyses**

The following criteria were used to select cathepsin-related genetic variants: (a) a $r^2$ LDmeasure of LD among SNPs < 0.001 within a 10,000 kb window and (b) a p value < the genome-wide significance level, i.e., $5 \times 10^{-6}$.

A genetic variant was considered a valid instrument if it fulfilled the following three core assumptions: (i) it was strongly associated with the exposure, (ii) it was independent of confounders between the exposure and outcome, and (iii) it exhibited no direct association with the outcome. The inverse variance-weighted (IVW) method was employed as the primary approach to estimate the overall effect size of an exposure on the outcome [34]. Statistical significance was assessed using adjusted p-values corrected for multiple comparisons; specifically, a p-value below 0.05/9 (accounting for nine cathepsins) was deemed statistically significant. Mendelian randomization analyses were conducted using the R packages TwoSampleMR [35] and MendelianRandomization [36].

Various statistical tests were conducted to evaluate the validity of the core instrumental variable assumptions. Cochran's Q test was employed to assess the heterogeneity among SNPs. The MR-Egger intercept[37] and the MR-PRESSO global test [38] were utilized to detect horizontal pleiotropy and identify outliers, respectively. The MR-PRESSO tests were executed using the R package MR-PRESSO [38].

Reverse MR analyses were conducted to evaluate reverse causality by treating EAC/BE as the exposure and cathepsin levels as the outcome. These analyses utilized the same GWAS datasets as those used in the forward MR analyses. Additionally, multivariable MR analyses were performed, incorporating multiple cathepsins as exposures to assess their respective effects after controlling for other cathepsin types.

**TWAS analysis**

For the TWAS analysis, the TWAS/FUSION method was applied to predict associations between the expression levels of specific cathepsins in esophageal tissue and the risk of EAC. The reference panel for esophageal tissue data was obtained from GTEx, version 8, as described in a previous study [39]. The tissue-specific expression levels were combined with EAC/BE GWAS traits to transform genetic variant-phenotype relationships into gene/transcript-phenotype relationships.

**scRNA-seq analysis**

The R Seurat package was employed to analyze the downloaded digital gene-cell matrix from the EAC samples [40]. Cells from different samples were filtered based on the criterion that each cell must detect at least 200 genes. A global scaling method was applied to normalize the gene expression matrices using the default scale factor, followed by a log(1+x) transformation. Louvain clustering was performed on the top 2000 highly variable genes, with clustering results visualized using uniform manifold approximation and projection (UMAP) [41]. DEGs were identified to annotate the clusters. The CellChat package was utilized for cell-cell communication analysis, and the igraph package was used to visualize the communication analysis results.

**Colocalization analysis**

The Coloc package [42] was used to perform colocalization analysis. Evidence for colocalization was evaluated using the posterior probability for Hypothesis 4 (PP4), which posits that two traits are associated and share the same causal variants. A PP4 threshold of 0.95 was used to confirm colocalization. Statistical analyses were conducted using R software version 4.1.1.

**Protein docking**

Molecular docking was conducted to predict potential interactions between two proteins. The protein structures were retrieved from the Protein Data Bank (PDB) and analyzed using the ZDOCK tool (https://zdock.wenglab.org/). Visualization of the docking results was performed using PyMol software.

**Immunohistochemistry**

The paraffin-embedded tissue samples were procured from the Department of Pathology at the First Hospital of Jilin University under ethical approval (AF-IRB-032-07). Immunohistochemistry (IHC) was conducted using an ultrasensitive SP IHC kit (Maixin, Fuzhou, China, KIT-9710). The paraffin sections were dewaxed in xylene, rehydrated in a graded alcohol series, and subject to antigen retrieval in boiling EDTA buffer. Endogenous peroxidase activity and nonspecific epitopes were blocked using reagents provided in the kit. The sections were incubated overnight at 4 °C with a 1:100 dilution of rabbit polyclonal antibodies against cathepsin B (Proteintech, 12216–1-AP). Biotinylated goat anti-rabbit secondary antibody, avidin, and biotinylated horseradish peroxidase were then applied sequentially. Hematoxylin was used for counterstaining. Images of the stained sections were captured with an Olympus microscope and quantified using the IHC profiler plugin for ImageJ. Statistical analyses were performed using GraphPad Prism software.

**Data availability**

The GWAS data of cathepsins and Macrophage were derived from the (https://gwas.mrcieu.ac.uk.). y statistics of EAC/BE were collected from https://www.ebi.ac.uk/gwas/. The scRNA seq data of EAC/BE store in the Gene Expression Omnibus (https://www.ncbi.nlm.nih.gov/geo/), access number is GSE173950. All packages for data analysis used in this study were open source in R software (version 4.1.1; R Development Core Team).

**References:**

1   Smyth, E. C. *et al.* Oesophageal cancer. *Nat Rev Dis Primers* **3**, 17048, doi:10.1038/nrdp.2017.48 (2017).

2   Cook, M. B., Chow, W. H. & Devesa, S. S. Oesophageal cancer incidence in the United States by race, sex, and histologic type, 1977-2005. *Br J Cancer* **101**, 855-859, doi:10.1038/sj.bjc.6605246 (2009).

3   Thrift, A. P. & Whiteman, D. C. The incidence of esophageal adenocarcinoma continues to rise: analysis of period and birth cohort effects on recent trends. *Ann Oncol* **23**, 3155-3162, doi:10.1093/annonc/mds181 (2012).

4   Salimian, K. J., Birkness-Gartman, J. & Waters, K. M. The path(ology) from reflux oesophagitis to Barrett oesophagus to oesophageal adenocarcinoma. *Pathology* **54**, 147-156, doi:10.1016/j.pathol.2021.08.006 (2022).

5   Naini, B. V., Souza, R. F. & Odze, R. D. Barrett's Esophagus: A Comprehensive and Contemporary Review for Pathologists. *Am J Surg Pathol* **40**, e45-66, doi:10.1097/pas.0000000000000598 (2016).

6   Chen, P. *et al.* Characterization of 500 Chinese patients with cervical esophageal cancer by clinicopathological and treatment outcomes. *Cancer Biol Med* **17**, 219-226, doi:10.20892/j.issn.2095-3941.2019.0268 (2020).

7   Man, S. M. & Kanneganti, T. D. Regulation of lysosomal dynamics and autophagy by CTSB/cathepsin B. *Autophagy* **12**, 2504-2505, doi:10.1080/15548627.2016.1239679 (2016).

8   Pišlar, A., Perišić Nanut, M. & Kos, J. Lysosomal cysteine peptidases - Molecules signaling tumor cell death and survival. *Semin Cancer Biol* **35**, 168-179, doi:10.1016/j.semcancer.2015.08.001 (2015).

9   Zheng, K. *et al.* Inhibition of autophagosome-lysosome fusion by ginsenoside Ro via the ESR2-NCF1-ROS pathway sensitizes esophageal cancer cells to 5-fluorouracil-induced cell death via the CHEK1-mediated DNA damage checkpoint. *Autophagy* **12**, 1593-1613, doi:10.1080/15548627.2016.1192751 (2016).

10  Breton, J. *et al.* Proteomic screening of a cell line model of esophageal carcinogenesis identifies cathepsin D and aldo-keto reductase 1C2 and 1B10 dysregulation in Barrett's esophagus and esophageal adenocarcinoma. *J Proteome Res* **7**, 1953-1962, doi:10.1021/pr7007835 (2008).

11  Fisher, O. M. *et al.* High Expression of Cathepsin E in Tissues but Not Blood of Patients with Barrett's Esophagus and Adenocarcinoma. *Ann Surg Oncol* **22**, 2431-2438, doi:10.1245/s10434-014-4155-y (2015).

12  Hughes, S. J. *et al.* A novel amplicon at 8p22-23 results in overexpression of cathepsin B in esophageal adenocarcinoma. *Proc Natl Acad Sci U S A* **95**, 12410-12415, doi:10.1073/pnas.95.21.12410 (1998).

13  Cheng, P. *et al.* Gene expression in rats with Barrett's esophagus and esophageal adenocarcinoma induced by gastroduodenoesophageal reflux. *World J Gastroenterol* **11**, 5117-5122, doi:10.3748/wjg.v11.i33.5117 (2005).

14  Fujiwara, N. & Kobayashi, K. Macrophages in inflammation. *Curr Drug Targets Inflamm*

*Allergy* **4**, 281-286, doi:10.2174/1568010054022024 (2005).

15   Shapouri-Moghaddam, A. *et al.* Macrophage plasticity, polarization, and function in health and disease. *J Cell Physiol* **233**, 6425-6440, doi:10.1002/jcp.26429 (2018).

16   Nicoletti, A. *et al.* The macrophage scavenger receptor type A directs modified proteins to antigen presentation. *Eur J Immunol* **29**, 512-521, doi:10.1002/(sici)1521-4141(199902)29:02<512::Aid-immu512>3.0.Co;2-y (1999).

17   Shirai, H. *et al.* Structure and function of type I and II macrophage scavenger receptors. *Mech Ageing Dev* **111**, 107-121, doi:10.1016/s0047-6374(99)00079-2 (1999).

18   Edgren, G., Adami, H. O., Weiderpass, E. & Nyrén, O. A global assessment of the oesophageal adenocarcinoma epidemic. *Gut* **62**, 1406-1414, doi:10.1136/gutjnl-2012-302412 (2013).

19   Buas, M. F. *et al.* Germline variation in inflammation-related pathways and risk of Barrett's oesophagus and oesophageal adenocarcinoma. *Gut* **66**, 1739-1747, doi:10.1136/gutjnl-2016-311622 (2017).

20   Casson, A. G. *et al.* Polymorphisms in DNA repair genes in the molecular pathogenesis of esophageal (Barrett) adenocarcinoma. *Carcinogenesis* **26**, 1536-1541, doi:10.1093/carcin/bgi115 (2005).

21   Ye, W. *et al.* The XPD 751Gln allele is associated with an increased risk for esophageal adenocarcinoma: a population-based case-control study in Sweden. *Carcinogenesis* **27**, 1835-1841, doi:10.1093/carcin/bgl017 (2006).

22   Bengsch, F. *et al.* Cell type-dependent pathogenic functions of overexpressed human cathepsin B in murine breast cancer progression. *Oncogene* **33**, 4474-4484, doi:10.1038/onc.2013.395 (2014).

23   Liu, J. *et al.* Cathepsin B and its interacting proteins, bikunin and TSRC1, correlate with TNF-induced apoptosis of ovarian cancer cells OV-90. *FEBS Lett* **580**, 245-250, doi:10.1016/j.febslet.2005.12.005 (2006).

24   Buck, M. R., Karustis, D. G., Day, N. A., Honn, K. V. & Sloane, B. F. Degradation of extracellular-matrix proteins by human cathepsin B from normal and tumour tissues. *Biochem J* **282 ( Pt 1)**, 273-278, doi:10.1042/bj2820273 (1992).

25   Meyer, N. *et al.* Autophagy activation, lipotoxicity and lysosomal membrane permeabilization synergize to promote pimozide- and loperamide-induced glioma cell death. *Autophagy* **17**, 3424-3443, doi:10.1080/15548627.2021.1874208 (2021).

26   Česen, M. H., Pegan, K., Spes, A. & Turk, B. Lysosomal pathways to cell death and their therapeutic applications. *Exp Cell Res* **318**, 1245-1251, doi:10.1016/j.yexcr.2012.03.005 (2012).

27   Ali, M. W. *et al.* A risk variant for Barrett's esophagus and esophageal adenocarcinoma at chr8p23.1 affects enhancer activity and implicates multiple gene targets. *Hum Mol Genet* **31**, 3975-3986, doi:10.1093/hmg/ddac141 (2022).

28   Onyishi, C. U. *et al.* Toll-like receptor 4 and macrophage scavenger receptor 1 crosstalk regulates phagocytosis of a fungal pathogen. *Nat Commun* **14**, 4895, doi:10.1038/s41467-023-40635-w (2023).

29   Sapkota, M., Kharbanda, K. K. & Wyatt, T. A. Malondialdehyde-Acetaldehyde-Adducted Surfactant Protein Alters Macrophage Functions Through Scavenger Receptor A. *Alcohol Clin Exp Res* **40**, 2563-2572, doi:10.1111/acer.13248 (2016).

30    Sun, B. B. *et al.* Genomic atlas of the human plasma proteome. *Nature* **558**, 73-79, doi:10.1038/s41586-018-0175-2 (2018).

31    Ahola-Olli, A. V. *et al.* Genome-wide Association Study Identifies 27 Loci Influencing Concentrations of Circulating Cytokines and Growth Factors. *Am J Hum Genet* **100**, 40-50, doi:10.1016/j.ajhg.2016.11.007 (2017).

32    Gharahkhani, P. *et al.* Genome-wide association studies in oesophageal adenocarcinoma and Barrett's oesophagus: a large-scale meta-analysis. *Lancet Oncol* **17**, 1363-1373, doi:10.1016/s1470-2045(16)30240-6 (2016).

33    The GTEx Consortium atlas of genetic regulatory effects across human tissues. *Science* **369**, 1318-1330, doi:10.1126/science.aaz1776 (2020).

34    Burgess, S., Butterworth, A. & Thompson, S. G. Mendelian randomization analysis with multiple genetic variants using summarized data. *Genet. Epidemiol.* **37**, 658-665, doi:10.1002/gepi.21758 (2013).

35    Hemani, G. *et al.* The MR-Base platform supports systematic causal inference across the human phenome. *eLife* **7**, doi:10.7554/eLife.34408 (2018).

36    Yavorska, O. O. & Burgess, S. MendelianRandomization: an R package for performing Mendelian randomization analyses using summarized data. *Int. J. Epidemiol.* **46**, 1734-1739, doi:10.1093/ije/dyx034 (2017).

37    Verbanck, M., Chen, C. Y., Neale, B. & Do, R. Detection of widespread horizontal pleiotropy in causal relationships inferred from Mendelian randomization between complex traits and diseases. *Nat. Genet.* **50**, 693-698, doi:10.1038/s41588-018-0099-7 (2018).

38    Verbanck, M., Chen, C. Y., Neale, B. & Do, R. Detection of widespread horizontal pleiotropy in causal relationships inferred from Mendelian randomization between complex traits and diseases. *Nat Genet* **50**, 693-698, doi:10.1038/s41588-018-0099-7 (2018).

39    Barbeira, A. N. *et al.* Exploring the phenotypic consequences of tissue specific gene expression variation inferred from GWAS summary statistics. *Nat Commun* **9**, 1825, doi:10.1038/s41467-018-03621-1 (2018).

40    Gribov, A. *et al.* SEURAT: visual analytics for the integrated analysis of microarray data. *BMC Med Genomics* **3**, 21, doi:10.1186/1755-8794-3-21 (2010).

41    Becht, E. *et al.* Dimensionality reduction for visualizing single-cell data using UMAP. *Nat Biotechnol*, doi:10.1038/nbt.4314 (2018).

42    Giambartolomei, C. *et al.* Bayesian test for colocalisation between pairs of genetic association studies using summary statistics. *PLoS Genet* **10**, e1004383, doi:10.1371/journal.pgen.1004383 (2014).

**Acknowledgements**

We thank Sagesci (www.sagesci.cn) for its linguistic assistance during the preparation of this manuscript. The summary figure was created by Scidraw and Biorender.

**Authors' contributions**

S.T and W.L conceived and designed the experiment; J.L and S.Y analyzed and verified the underlying data; J.L and S.T wrote the original manuscript. M.T and X.G involved in data interpretation. X.M supported pathological section preparation and pathological interpretation. All authors have read and approved the final version of the manuscript.

**Funding**

This study was funded by Jilin Province medical and health talents special (JLSWSRCZX2023-35). This study was funded by Jilin Provincial Science and Technology Development Plan Project (20220204115YY) and Jilin Provincial Science and Technology Development Plan Project, Natural Science Foundation of Jilin Province (YDZJ202301ZYTS007 and YDZJ202201ZYTS121). The funding body had no role in the design of the study and collection, analysis and interpretation of data and in writing the manuscript.

**Competing interests**

All other authors declare no competing financial interests. All authors approved the final version of the manuscript.

**Figure legends:**

**Figure 1. Forest plot of the univariable Mendelian randomization analysis between nine cathepsins and esophageal adenocarcinoma/Barrett's esophagus (EAC/BE) risk.** The inverse variance-weighted

(IVW) method was used to determine the causal relationships between the nine cathepsin types and EAC/BE and between EAC and BE. The statistically significant results were highlighted in red and the suggestively significant results were highlighted in green. The error bars corresponded to 95% confidence intervals.

**Figure 2. Forest plot of the multivariable Mendelian randomization analysis between cathepsins and esophageal adenocarcinoma/Barrett's esophagus (EAC/BE) risk.** The inverse variance-weighted (IVW) method was used to determine the causal associations between nine cathepsins (cathepsin B, E, F, G, H, L2, O, S, and Z) and EAC/BE and between EAC and BE. The statistically significant results were highlighted in red, and the error bars represented 95% confidence intervals.

**Figure 3. Immunohistochemical staining of CTSB in normal and EAC tissues.** (a) The expression of CTSB in normal esophageal tissues and EAC tissues. (b) The statistical analysis results of the immunohistochemical analysis. Bars represent means, and error bars represent 95% confidence intervals; * represents $p < 0.05$.

**Figure 4. Visualization of single-cell RNA sequencing data analysis for esophageal adenocarcinoma (EAC) tissues.** (a) UMAP plot of 43,349 single cells in EAC tissues color-coded based on the eight major lineages. (b) Dot plot of the mean expression of the canonical marker genes for the eight major EAC lineages. (c) UMAP visualization of the primary distribution of cathepsin B (CTSB) in myeloid cells of esophageal adenocarcinoma patients. (d) UMAP plot for the myeloid cells in patients with EAC color-coded based on the three major lineages. (d) Dot plot of the mean expression of the canonical marker genes for macrophages, monocytes, and dendritic cells. (f) UMAP visualization of the primary distribution of cathepsin B (CTSB) in the macrophages of patients with EAC.

**Figure 5. The results of the cell communication analysis of CTSB+ macrophages and CTSB-

**macrophages.** (a) Network diagrams of the number and strength of cell-cell interactions. (b) Heatmap of the overall outgoing and incoming signaling patterns among the cell subpopulations of EAC patients. (c) Network diagrams of the differences in the number of cell-cell interactions between CTSB+_ macrophages/CTSB-_ macrophages and other cell subpopulations of EAC patients. (d) Network diagrams of the strength of the cell-cell interactions between CTSB+_ macrophages/CTSB-_ macrophages and other cell subpopulations of EAC patients. (e) Dot plots showing the ligand-receptor pair interactions between CTSB+_ macrophages/CTSB-_ macrophages and the other cell subpopulations of EAC patients.

**Figure 6. Molecular docking diagram of CTSB and MSR.** The molecular structure of MSR is presented in green, and that of CTSB is presented in blue. The hydrogen bonds between the CTSB-MSR complex are represented as yellow sticks, and the related residues are labeled with one letter.

**Table 1:** Transcriptome-wide association study (TWAS) analysis to determine the gene expression levels of cathepsin B using normal esophageal tissue as the reference

| Cohorts | Z score | P value | $R^2$ |
|---|---|---|---|
| esophageal adenocarcinoma/Barrett's esophagus | -4.575 | $4.76 \times 10^{-6}$ | 0.3 |
| esophageal adenocarcinoma | -3.088 | 0.002 | 0.3 |
| Barrett's esophagus | -3.791 | $1.50 \times 10^{-4}$ | 0.3 |

**Supplementary Files**

Supplementary Table 1. Sensitivity analysis results for univariable MR analyses between cathepsins and EAC/BE, EAC, and BE.

Supplementary Table 2. Reverse MR analysis results between the cathepsin family and EAC/BE.

Supplementary Table 3. Two-sample MR results between cathepsin B and macrophage receptors.

Supplementary Figure 1. Expression levels of CTSB in macrophages across tumor size stages (T stage) for the patients with EAC.

Supplementary Figure 2. UMAP plot and dot plot used to annotate CTSB+ macrophage and CTSB- macrophage subpopulations.